\documentstyle[12pt]{article}

\newcommand{\half}{\frac{1}{2}}
\newcommand{\M}{\cite{Mazur}}
\newcommand{\R}{\cite{Ren}}

\newcommand{\D}{\cite{Deser}}
\newcommand{\Lo}{\cite{Lousto}}
\newcommand{\MLR}{\cite{Mazur,Lousto,Ren}}
\newcommand{\ML}{\cite{Mazur,Lousto}}
\newcommand{\A}{\cite{Aharonov}}
\begin{document}
\begin{titlepage}
\setcounter{page}{1}
\title{\bf Gravitational scattering on a global monopole reexamined}
\author{Waldemar Puszkarz\thanks{Electronic address: puszkarz@math.sc.edu}
\\
\small{\it  Department of Physics and Astronomy,}
\\
\small{\it University of South Carolina,}
\\
\small{\it Columbia, SC 29208}}
\date{\small (5 November 1995)}
\maketitle
\begin{abstract}
We critically reexamine the gravitational scattering of scalar particles
on a global monopole studied recently. The original investigation of
Mazur and Papavassiliou is
extended by considering different couplings of the scalar field to the
space-time curvature and by varying the dimension of the space-time where
the Klein-Gordon (KG) field lives. A universal behavior of the leading term
in the scattering amplitude as a function of this dimension is revealed.
\vskip 0.5cm
\vskip 3cm
\flushright{Submitted to Physical Review D}
\end{abstract}

\end{titlepage}

\section{Introduction}
Recently, the gravitational scattering on a global monopole has been
studied \MLR. In the most recent paper \R, the scattering of fermions
has been investigated and the correct scattering amplitude for it derived.
The other papers \ML\, deal with scalar particles in a monopole
background. Their analysis of the scattering of these particles on the
global monopole leads to equations describing the total scattering
cross-section for this process. However, a brief examination of those
equations suffices to notice that they
differ, having in common a serious pathology: in the limit
corresponding to the Minkowski space-time they result in infinite
total cross-sections for the process under consideration. Thus, one can
rightly suspect that this very unphysical result is due to some errors
in the derivation of the discussed formulas rather than an intrinsic
pathology of the problem. To show that it is indeed the case is in part
the goal of the present paper. Before we embark on this, let us
mention key errors in \ML. In the approach of Mazur and Papavassiliou
\M, the optical theorem is used to obtain the expressions for the total
scattering cross-section. Since, as shown below, this
theorem is not valid for the case under study,
its application renders formulas describing the total
cross-section (equations (26) and (28) in [1]) incorrect. In the paper by
Lousto, a simple error in the calculation of an integral is made (formulas
(27), (28), and (25) in \Lo\,), which nevertheless brings about dramatic
consequences. The present report is organized as follows. Since its major
purpose is to reexamine and elucidate points that
lead to wrong results in the above referred papers, we perform this task
in the very next section where we also find out
how the inclusion of interaction between the scalar field and the curvature
of the global monopole space-time affects the amplitude for the process
under consideration. We extend the
original investigations \ML\, in Section 3. Here, we demonstrate that
the leading term
in the amplitude can be obtained as a special case of the scattering on
the global monopole in an arbitrary number of dimensions $D=1+d$ of the
underlying space-time with $d>1$. In the conclusions we
summarize the main results of the work presented. Following this, two
appendices gather explicit derivations
of some relevant formulas, which we did not find suitable to include
in the main body of the paper.

\section{Critique and Correction of Previous Results}

For the sake of simplicity, in what follows, we will study massless
scalar particles in the global monopole background. We will adopt a system
of units in which $c=G=\hbar=1$.

Let us start from the Lagrangian
$$
L={\half} g^{\alpha\beta}\nabla_{\alpha}\nabla_{\beta}\Psi +\xi
R{\Psi}^2, \eqno(1)
$$
where $\nabla_{\alpha}$ is a covariant derivative with respect to the
metric $g_{\alpha\beta}$
and $R$ stands for the scalar curvature of the space-time where the scalar
field $\Psi$ propagates. The parameter $\xi$  is a coupling constant.

The equation of motion derived from (1) reads
$$
\left(\nabla^2-\xi R\right)\Psi=0, \eqno(2)
$$
where
$$
\nabla^2={\frac{1}{\sqrt{-g}}}\partial_{\alpha}\left(
\sqrt{-g}\,\partial_{\beta}g^{\alpha\beta}\right)
$$
and $g=det(g_{\alpha\beta})$.

We will first consider the minimal coupling case, i.e., with $\xi=0$.
Since the metric of the global monopole
$$
ds^2=-dt^2+dr^2+b^2r^2(d\theta^2+\sin^2\theta d\phi^2) \eqno (3)
$$
is static, the stationary solutions to (2) can be assumed as $\Psi(\vec r,t)=
\Phi(\vec r)e^{-iEt}$. (See [4] for the derivation of the metric and [1]
for a list of references.) Upon defining $k^2=E^2$, the KG equation reduces to
$$
\left[{\frac{1}{r^2}}{\frac{\partial}{\partial
r}}\left(r^2{\frac{\partial}{\partial
r}}\right)-{\frac{L^2}{b^2r^2}}+k^2\right] \Phi(\vec r)=0, \eqno (4)
$$
where $L^2$ is the Laplacian on a two-dimensional sphere and ${k^2>0}$
as we are interested in the scattered states only.

Due to the spherical symmetry of the problem under study, as a complete
set of egeinfunctions on $S^2$ one can employ the Legendre polynomials
$P_l(\cos\theta)$. Let us recall that they satisfy
$$
L^2P_l(\cos\theta)=l(l+1)P_l(\cos\theta). \eqno (5)
$$
Looking for a solution to (2) as a series
$\sum\limits_{l=0}^{\infty}a_lR_l(r)P_l(\cos\theta)$, brings us, upon
the separation of variables,  to the following
equation for the radial wave function
$$
R''+{\frac{2}{r}}R'+\left(k^2-{\frac{l(l+1)}{b^2r^2}}\right)R=0,
\eqno (6)
$$
the primes denoting the differentiation with respect to $r$.
By demanding that $R$ is regular at the origin, one can single out the unique
solution $R_l(r)=r^{-1/2}J_{\nu(l)}(kr)$,
where $\nu(l)=b^{-1}\sqrt{(l+1/2)^2-(1-b^2)/4}$
{}~(see [1] for more details).

The general scattering solution is a superposition of the incoming wave
function $\Phi_{in}=e^{ikz}=e^{ikr\cos\theta}$ and the scattered one $\Phi_{
scat}$
$$
\Phi=\Phi_{in}+\Phi_{scat}, \eqno (7)
$$
where $\Phi_{scat}=e^{ikr}{{f(\theta)}\over r}$ as $r\rightarrow\infty$.
We are primarily interested in the function $f(\theta)$, the scattering
amplitude that contains information on the differential cross-section of
the scattering process. Using the partial wave analysis, one arrives at
the following formula for this amplitude [see Eq.(B13)],
$$
f(\theta)={\frac{1}
{2ik}}\sum_{l=0}^{\infty}(2l+1)(e^{2i\delta_l}-1)P_l(\cos
\theta), \eqno (8)
$$
where a phase shift $\delta_l$ is a function of $b$
$$
\delta_l=\delta_l(b)={\frac{\pi}{2}}\left(l+1/2-\nu(l)\right)=
{\frac{\pi}{2}}\left(l+1/2
-b^{-1}\sqrt{(l+1/2)^2-(1-b^2)/4}\right) .\eqno (9)
$$

It is here that our analysis departs significantly from the one in [1].
First of all, the sum
$$
\Sigma_{0}=\sum_{l=0}^{\infty}(2l+1)P_l(\cos\theta), \eqno (10)
$$
as shown in Appendix A, produces a Dirac-delta term which was overlooked
in [1]. Besides, as we will shortly see, even if we neglect this term,
the remaining part of the amplitude leads to a differential scattering
cross-section
$$
{\frac{d\sigma}{d\Omega}}=\vert f(\theta)\vert^2 \eqno (11)
$$
that being strongly divergent for some values of $\theta$, results in an
infinite total scattering cross-section $\sigma$  for our process. The
optical theorem,
$$
\sigma= {\frac{4\pi}{k}} Imf(0), \eqno (12)
$$
usually invoked to relate $f(\theta)$  to $\sigma$,  even if naively
valid in the sense that both sides of it are infinite, does not provide
any meaningful
information on the scattering process under consideration. Moreover,
the optical theorem is formally (naively) valid only due to the presence
of the delta term $\Sigma_{0}$ that was omitted in [1].

We will now proceed to compute
$$
\Sigma_{r}=\sum_{l=0}^{\infty}(2l+1)e^{2i\delta_l}P_l(\cos\theta) \eqno (13)
$$
by employing the method of [1].
Since $e^{2i\delta_l}=e^{i\pi\alpha z}(1+i{\pi}a^2/2bz+
O(z^{-3}))$,  where  $\alpha=1-b^{-1}$,
$a^2=(1-b^2)/4$, and  $z=l+1/2$, if we use $a^2$ as an expansion parameter,
$\Sigma_r $
can be well approximated by the first two terms in this expansion, i.e.,
$\Sigma_r=\Sigma_1+\Sigma_2$, where
$$
\Sigma_1=2\sum_{l=0}^{\infty}z(l)e^{i\pi\alpha z(l)}P_l(\cos\theta)=
{\frac{2}{i\pi}}{\frac{d}{d\alpha}}\sum_{l=0}^{\infty}e^{i\pi\alpha
z(l)}P_l(\cos\theta)=\frac{-2i\sin\pi\alpha}
{\left[\,2(\cos\pi\alpha-\cos\theta)\, \right]^{3/2}}, \eqno (14)
$$
and
$$
\Sigma_2=\frac{i{\pi}a^2}{b}\sum_{l=0}^{\infty}e^{i\pi\alpha
z(l)}P_l(\cos\theta)
=\frac{i{\pi}a^2b^{-1}}{\left[\,2(\cos\pi\alpha-\cos\theta)\,\right]^{1/2}}.
\eqno (15)
$$
To show that $\sum\limits_{l=0}^{\infty}e^{i\pi\alpha
z(l)}P_l(\cos\theta)=
\left[2(\cos\pi\alpha-\cos\theta)\right]^{-1/2}$,
one makes use of the generating function for the Legendre polynomials
$F(h,\theta)=\sum\limits_{l=0}^{\infty}h^lP_l(\cos\theta)=\left
(1-2h\cos\theta+h^2\right)^{-1/2}.$
Therefore, in our approximation the scattering amplitude for
$\theta<\pi\alpha$  and $\theta>\pi\alpha$ is
$$
{f_{-}(\theta)}=-{\frac{i}{k}}
\left[ \delta(1-\cos\theta)+{\frac{1}
{2\sqrt{2(\cos\theta-\cos\pi\alpha)}}}
\left( {\frac{\pi a^2}{b}}+{\frac{\sin\pi\alpha}
{\cos\theta-\cos\pi\alpha}}\right) \right], \eqno (16a)
$$
and
$$
{f_{+}(\theta)}={\frac{1}
{2k\sqrt{2(\cos\pi\alpha-\cos\theta)}}}\left[\frac{\pi a^2}{b}-
\frac{\sin\pi\alpha}{\cos\pi\alpha-\cos\theta}\right], \eqno (16b)
$$
correspondingly.
As seen from the last equations, the differential cross-section is singular
for $\theta=0,~\pi\alpha$. It seems also to be vanishing for $\theta$ such
that $|\cos\pi\alpha-\cos\theta|=b\sin\pi\alpha/{\pi a^2}$.
However, this effect is obviously an artifact of the approximation used and
if the omitted terms are present, the scattering cross-section is not likely
to vanish, at least not for the angles satisfying the above condition.

For $\theta\ne 0$, the differential cross-section simplifies to
$$
{\left(\frac{d\sigma}{d\Omega}\right)}_{\mp}=\frac{\sin^2\pi\alpha}
{64k^2\left|\sin^3\frac{\omega}{2}\sin^3{(\frac{\omega}{2}+\pi\alpha)}\right|}
\left[1-
\frac{2{\pi}a^2\sin\frac{\omega}{2}\sin(\frac{\omega}{2}+\pi\alpha)}
{b\sin\pi\alpha}\right]^2,  \eqno (17)
$$
where $\omega=\theta- \pi\alpha$.
Clearly, the main contribution to the cross-section comes
from the vicinity of the ring $\theta=\pi\alpha$, which corresponds to
the limit $\omega\rightarrow 0$ in Eq.(17) and should physically be
understood as the condition $\omega\ll \pi\alpha$. It is in this limit
that the
singularity of the scattering cross-section for a non-forward scattering
is displayed in its full transparency. In a good approximation the
differential cross-section is given here by
$$
\frac{d\sigma}{d\Omega}\approx\frac{1}{64k^2|\sin\pi\alpha
\sin^3\omega/2|}. \eqno (18a)
$$
As seen from this formula, the singularity so strong is bound to yield
an infinite total cross-section even if the delta term is discarded.
In the regime $\alpha\ll 1$, which seems to be the most physically plausible,
the leading approximation to $\frac{d\sigma}{d\Omega}$ for $\omega$
not so close to $0$ behaves as
$$
\frac{d\sigma}{d\Omega}\approx
\frac{(\pi\alpha)^2}{64k^2\sin^6\omega/2}\left(1+\sin^2\omega/2
\right)^2 . \eqno (18b)
$$
It is through Eqs.(17) and (18) that the physical signature of the monopole
on the scattering of scalar particles is most prominently exhibited.

It is a good point to address some results of \Lo\, pertinent to our study.
Contrary to Eq. (28) in there, the total scattering cross-section
is not finite as shown above. The roots of this error are in the
omission of
the absolute value brackets in Eq. (25) of the discussed paper and the
subsequent application of the Cauchy principal value method to calculate
an integral whose integrand does not surrender to the trick: it does not
change sign at the singular point with the consequence of infinities
adding up instead of cancelling out. The question whether the total
scattering cross-section is to be regularized will not be
considered in this report. Although in the addressed paper this issue
seems to be brought up by the error, one should not neglect it. We limit
ourselves to two comments only.
On the one hand, it is not a new problem: a similar singularity occurs in
the Rutherford formula for the Coulombic scattering as it does in the
celebrated Aharonov-Bohm scattering \A\,. In both cases the total
cross-section is infinite.
On the other hand, it is clear that if one wants  to make any reasonable
use of the total cross-section in situations like the discussed ones,
a regularization cannot be avoided.

Let us now consider the case $\xi\ne 0$. Since for our metric the
scalar curvature is $R=-2(1-1/{b^2})r^{-2}$ ~(see [1]), Eq.(4) is
now replaced by
$$
\left[\frac{1}{r^2}\frac{\partial}{\partial r}(r^2\frac{\partial}
{\partial r})-\frac{L^2}{b^2r^2}-\frac{2\xi(1-1/{b^2})}{r^2}+k^2 \right]
\Phi(\vec r)=0 \eqno (19)
$$
and the radial part of it can be brought to
$$
R''+\frac{2}{r}
R'+\left[\frac{1}{r^2}\left(l(l+1)/{b^2}+2\xi(1-1/{b^2})\right)
 +k^2 \right]R=0 \eqno (20)
$$
whose solution is $ R_{\nu(l,\xi)}=r^{-1/2}J_{\nu(l,\xi)}(kr)$,
where $\nu(l,\xi)=b^{-1}\sqrt{(l+1/2)^2-(1-b^2)(1+8\xi)/4}$.
We see that the only way $\xi$ affects our previous solution is by the
change of the parameter $a^2$, namely $a^2\rightarrow a'^2=a^2(1+8\xi)$.
For $\xi=-1/8$, $a'^2=0$, causing formulas (16) and (17) to
simplify considerably. For the massless, conformally coupled scalar field
$\xi=1/6$ and $a'^2=7(1-b^2)/12$ is slightly more than
twice as large as for the massless, non-coupled field. (Similar observations
in congruence with ours are made in \Lo.)

\section{Higher Dimensional Scattering}

We will now establish a universal behavior of the leading term in the
differential scattering cross-section for the scattering on the global
monopole in an arbitrary number of dimensions.

To this end, let us consider the metric of the global monopole in $D=1+d$
dimensions ($d\ge 2$)
$$
ds^2=-dt^2+dr^2+b^2r^2d\Omega_{d-1}^2, \eqno (21)
$$
where $d\Omega_{d-1}^2$ is the metric on a $(d-1)$-\, dimensional sphere and
$b^2$ reperesents the monopole defect which in (1+2)- and (1+3)-\,dimensional
spacetimes is the angular and the solid angle defect, respectively. For $d>3$
one has to do with defects that result in deficits on spheres of higher
dimensions. In what follows, we will assume $d>2$.

For simplicity, let us limit ourselves to the minimal coupling case so that
the KG equation for $\Psi=e^{iEt}\Phi(\vec r)$ reduces to
$$
\left[\frac{1}{r^d}\frac{\partial}{\partial r}(r^d\frac{\partial}{\partial r})
-\frac{L_{d-1}^2}{b^2r^2}+k^2\right]\Phi(\vec r)=0 \eqno (22)
$$
[see the derivation of Eq.(4)], where $L_{d-1}^2$  is the Laplacian on the
$(d-1)$-~dimensional sphere. The eigenfunctions $Y$ of $L_{d-1}^2$ satisfy
the equation
$$
L_{d-1}^2Y=l(l+d-2)Y \eqno (23)
$$
and in the case of  spherical symmetry are known as the Gegenbauer
polynomials $G_l^s(\cos\gamma)$, where $s=\frac{d-2}{2}$ and $\gamma$
is an angle between two arbitrarily chosen directions from the center
of symmetry.
In what follows, we will treat $\gamma$ as the angle between an incoming and
a scattered wave.

We will seek the solution for $\Phi(\vec r)$ as a series
$$
\Phi(\vec r)=\sum_{l=0}^{\infty}a_lR_l^s(r)G_l^s(\cos\gamma) \eqno (24)
$$
the radial part of which satisfies
$$
\frac{1}{r^d}\frac{d}{dr}(r^d\frac{d}{dr})R_l^s+\left(k^2-\frac{l(l+d-2)}
{b^2r^2}\right) R_l^s=0. \eqno (25)
$$
By substituting $R_l^s=r^{-{(d-2)}/2}R_l$ one obtains from (25)
$$
R''_l+\frac{1}{r}R'_l+\left(k^2-\frac{\nu^2(l,s)}{r^2}\right)R_l=0,
\eqno (26)
$$
where $\nu^2(l,s)=\frac{(l+s)^2+s^2(1-b^2)}{b^2}$. The solution
regular at $r=0$ is $R_l^s(r)=r^{-s}J_{\nu(l,s)}(kr)$.

The amplitude of the scattering cross-section, as found in Appendix B, is
$$
f(\gamma)=\frac{C(d)}{k^{{(d-1)}/2}}\sum_{l=0}^{\infty}(l+s)(e^{2i
\delta_l(s)}-1)G_l^s(\cos\gamma)=\frac{C(d)}{k^{{(d-1)}/2}}
\left(\Sigma_0^s +\Sigma_r^s\right), \eqno (27)
$$
where $C(d)$ is constant for a fixed $d$ and
$\Sigma_0^s=\sum\limits_{l=0}^{\infty}(l+s)G_l^s(\cos\gamma)$ is zero (unless
$\gamma=0$ when it is infinite as a Dirac delta as shown in Appendix A) and
will be omitted from
further considerations. The other term, $\Sigma_r^s=\sum\limits_{l=0}^{\infty}
(l+s)e^{2i\delta_l(s)}G_l^s(\cos\gamma)$, will be studied in a greater detail
for it is the one that contains the leading term of $f(\gamma)$.
Now,
$$
\delta_l(s)=\frac{\pi}{2}\left(l+s -\nu(l,s)\right)=\frac{\pi}{2}\left[z-
\frac{z}{b} \sqrt {1-{a^2}/{z^2}}\,\right]=\frac{\pi}{2}(\alpha z
+{a^2}/2bz+ O(z^{-3})), \eqno (28)
$$
where $\alpha=1-b^2$, $a^2=s^2(1-b^2)$, and $z=l+s$.
We are now ready to find the leading term $lt$ in $\Sigma_r^s$,
$$
\Sigma_r^s=\sum_{l=0}^{\infty}z(l)e^{i\pi\alpha z(l)}(1+{i\pi a^2}/2bz
+ O(z^{-2}))G_l^s(\cos\gamma), \eqno (29)
$$
that is,
$$
lt=\sum_{l=0}^{\infty}z(l)e^{i\pi\alpha z(l)}G_l^s(\cos\gamma)=\frac{1}
{i\pi}\frac{d}{d\alpha}h^s(\gamma,\alpha), \eqno (30)
$$
where
$$
h^s(\gamma, \alpha)=\sum_{l=0}^{\infty}e^{i\pi\alpha z(l)}G_l^s(\cos\gamma)=
\left[\,2(\cos\pi\alpha-\cos\gamma)\,\right]^{-s}. \eqno (31)
$$
The last formula can be worked out by performing straightforward
manipulations on the generating function for the Gegenbauer polynomials
$$
G(x,t)=\sum_{l=0}^{\infty}t^lG_l^s(x)=(1-2xt+t^2)^{-s}. \eqno (32)
$$
Finally,
$$
lt=\frac{-is\sin\pi\alpha}{2^s(\cos\pi\alpha-\cos\gamma)^{s+1}}
\eqno (33)
$$
and the leading term of $f(\gamma)$ in its complete form reads
$$
flt=\frac{-i(d-2)\Gamma(\frac{d-2}{2})e^{-i\frac{\pi}{4}(d-1)}\sin\pi\alpha}
{2^{\frac{3}{2}}\sqrt{\pi}k^{\frac{d-1}{2}}(\cos\pi\alpha-\cos\gamma)^
{\frac{d}{2}}} \eqno (34)
$$
The last formula is valid for $d>2$ and $\gamma>\pi\alpha$. It is easy to
analytically extend it to $\gamma<\pi\alpha$, but since this changes only
the exponent, the final result [Eq. (36)] remains unaffected.
For $d=2$ as shown in \D
$$
f(\gamma)\propto \frac{\sin\pi\alpha}
{k^{\half}(\cos\pi\alpha-\cos\gamma)}. \eqno (35)
$$
As opposed to $d>2$, the scattering amplitude on the monopole in
$(1+2)$-~dimensional space-time is known in its complete exact form.
Therefore we have shown that in the first approximation
$$
f(\gamma)\propto \frac{\sin\pi\alpha} {k^{\frac{d-1}{2}}(\cos\pi\alpha-
\cos\gamma)^{\frac{d}{2}}}, \eqno (36)
$$
thus exhibiting a universal behavior in the sense that it can be
described by a single unique formula which as a function of $d$ applies
to all dimensions $d>1$.

\section{Conclusions}

We have shown that the optical theorem as employed in [1]
for the gravitational scattering on the global monopole is not valid, thus
producing the wrong total cross-section for the process under study.
The situation here is similar to the Coulomb scattering where the
total cross-section is divergent due to the long-range nature of this
interaction and is characteristic to the scattering on the global monopole
in any number of dimensions in which the global monopoles are conceivable.
For the $(1+2)$-dimensional monopole it was first noticed in \D. It is
the cone-like structure of these space-times that provides the long-range
interaction thereby making even high angular momenta contribute in a
non-negligible manner to the total scattering cross-section.
Although the appearance of the delta term in the scattering amplitude is
the main obstruction in a meaningful application of the optical theorem,
the theorem would not be valid even if this term were left out.
Furthermore, we have provided the
resolution to the problem of the infinite total cross-section arising
in the Minkowski space-time limit reported in \Lo\,. This effect is
completely spurious as caused by erroneous calculations.
We have also demonstrated that the coupling of the scalar field to a non-zero
curvature of the monopole space-time can affect the scattering amplitude
leading in some instances to its simplification. A nice feature of the
amplitude is that its leading term is given by a single universal formula
valid for all space-time dimensions that allow for the existence of global
monopoles. In all of them, the total scattering cross-section is an
ill-defined quantity and, unless a reasonable regularization is proposed,
one should not invoke it to describe the scattering process discussed
throughout this paper.

\section*{Acknowledgments}

I would like to thank Professor Pawel O. Mazur
for introducing me to the physics of monopoles and  his encouragment to
write up the results of this work. I am particularly indebted to him for
the critical reading of the manuscript, which  helped clarify a few points
and correct some formulas. I am also beholden to Carlos O. Lousto for
pointing me to his paper. This work was partially supported by the NSF grant
No.13020 F167.

\section*{Appendix A}

We will show here that $\Sigma_0^s$ is equal to some Dirac-delta term.
Using the generating function for the Gegenbauer polynomials
$$
G(h,\gamma)=\sum_{l=0}^{\infty}h^lG_l^s(\cos\gamma)=(1-2h\cos\gamma+h^2)^{-s},
\eqno (A1)
$$
it is straightforward to see that
$$
\Sigma(h,x)= sG+h\frac{\partial G}{\partial h}
=\sum_{l=0}^{\infty}(l+s)h^lG_l^s
(\cos\gamma)=\frac{s(1-h^2)}{(1-2hx+h^2)^{s+1}}, \eqno (A2)
$$
where $x=\cos\gamma$, and $s=\frac{d-2}{2}$. Now,
$$
\Sigma_0^s=\sum_{l=0}^{\infty}(l+s)G_l^s(\cos\gamma)=\lim_{h\to 1}\Sigma(h,x)
$$
which equals $\infty$ if $x=1$ or $0$ otherwise.
This clearly demonstrates that $\Sigma_0^s$ is equal to a Dirac delta
term. For the $(1+3)$-~dimensional space-time $\Sigma_0^s={1\over 2}\Sigma_0$,
so $\Sigma_0=\sum_{l=0}^{\infty}(2l+1)P_l(\cos\theta)$ must be
proportional to some Dirac delta as well. To establish the coefficient of
proportionality, we will employ the orthogonality relation for the Legendre
polynomials
$$
\int_{-1}^{1}P_l(x)P_{l'}(x)\, dx=\frac{2}{2l+1}\delta_{ll'}. \eqno
(A3)
$$
By multiplying $\Sigma_0$ by $P_{l'}(x)$   and integrating over $x$, one
obtains $2$ due to (A3).
Therefore one is lead to conclude that $\Sigma_0=2\delta(1-\cos\theta)$.

\section*{Appendix B}

Below we present the phase-shift method, also known as the partial wave
analysis, applied to a spherically symmetric scattering in $D=1+d$ ($d>2$)
space-time dimensions.

In $D$ dimensions, the stationary Schr\"{o}dinger equation describing a single
freely propagating particle of mass $m$ and energy $E$ reads
$$
({\nabla_D}^2 + k^2)\Psi(\vec r)=0, \eqno (B1)
$$
where $k^2=2mE$ and
${\nabla_D}^2=\frac{1}{r^d}\frac{\partial}{\partial r}
 \left(r^d\frac{\partial}{\partial r}\right) -
\frac{L_{d-1}^2}{r^2}$, ~$L_{d-1}^2$ being the Laplacian on a
$(d-1)$~-dimensional sphere.
By separating variables $\Psi(\vec r)=R(r)Y(\gamma)$, one obtains
$$
L_{d-1}^2Y_l^d(\gamma)=l(l+d-2)Y_l^d, \eqno (B2)
$$
$$
R''(r)+  \frac{d}{r}R'(r) + \left(k^2-\frac{l(l+d-2)}{r^2}\right)R(r)=0,
\eqno (B3)
$$
where $\gamma$ is an angle between two arbitrary directions from the center
of symmetry which in our case is the center of scattering. In what follows,
we will think of $\gamma$ as the angle between an incoming and a scattered
wave.
The solutions to (B2) in terms of functions of $\cos\gamma$ are called the
Gegenbauer polynomials. We will use them from now on employing the notation
$Y_l^d(\gamma)=G_l^s(\cos\gamma)$, where $s=\frac{d-2}{2}$.

Let us now work out the solutions to (B3). We will denote them by $R_l^s$ by
analogy to $G_l^s$. Upon substituting $R_l^s=r^{-s}R_l$,  Eq. (B3)
reduces to
$$
R''_l+\frac{1}{r} R'_l+ \left(k^2-\frac{(l+s)^2}{r^2}\right)R_l=0,
\eqno (B4)
$$
that is to the well known Bessel equation. Being interested in the
solutions regular at $r=0$, we choose $R_l^s(r)=r^{-s}J_{l+s}(kr)$.
Once we know the solutions to (B3), the most complete solution to (B1)
can be found as a series $\Psi(\vec r)=
\sum\limits_{l=0}^{\infty}a_lR_l^s(r)G_l^s(\cos\gamma)$.
Knowing that the asymptotic behavior of $J_{\pm \nu}(y)$ for $y\rightarrow
\infty$ is
$$
J_{\pm \nu}(y)\sim \sqrt{\frac{2}{\pi y}}\cos\left(
y \mp \frac{\pi\nu }{2}-\frac{\pi}{4}\right), \eqno (B5)
$$
one obtains that for $r\rightarrow \infty$
$$
\Psi(\vec
r)\sim \sum_{l=0}^{\infty}b_l\cos\left(kr-\frac{\pi}{2}l-\frac{\pi}{4}
(d-1)\right)G_l^s(\cos\gamma)\frac{1}{r^{{(d-1)}/2}}. \eqno (B6)
$$
In the presence of a potential $V(r)$ modyfying Eq.(B1) through its impact
on $k^2$, one should expect phase shifts in the asymptotic form of
$\Psi(\vec r)$
$$
{\Psi}_{\infty}^V(\vec
r)=\sum_{l=0}^{\infty}c_l\cos\left(kr-\frac{\pi}{2}l
-\frac{\pi}{4}(d-1)+{\delta}_l(s)\right)G_l^s(\cos\gamma)
\frac{1}{r^{{(d-1)}/2}}. \eqno (B7)
$$
On the other hand, one can compose ${\Psi}^V(\vec r)$ of the incoming wave
$e^{ikr}$ and the scattered one
$$
{\Psi}^V(\vec r)=e^{ikr}+\frac{f(\gamma)}{r^{\frac{d-1}{2}}},
\eqno (B8)
$$
where $f(\gamma)$ is  the scattering amplitude and the incoming wave
propagates along the $z$~-axis. It is this axis with respect to which the
angle
$\gamma$ is measured. In an arbitrary number of dimensions, $e^{ ikz}$
can be expanded as
$$
e^{ikz}=2^s\Gamma(s)\sum_{l=0}^{\infty}(s+l)i^l
\frac{J_{(l+s)}(kr)}{(kr)^s}G_l^s(\cos\gamma) \eqno (B9).
$$
Using (B9), (B5), and (B8), the asymptotics of ${\Psi}^V
(\vec r)$ can be expressed  in terms of $f(\gamma)$ as

$$
{\Psi}_{\infty}^V(\vec r)=\frac{1}{r^{\frac{d-1}{2}}}\left[f(\gamma)e^{ikr}+
\frac{2^{\frac{d-3}{2}}\Gamma(s)}{\sqrt{\pi}k^{\frac{d-1}{2}}}
\sum_{l=0}^{\infty}
(2l+d-2)i^l\cos(kr-\frac{\pi}{2}l-\frac{\pi}{4}(d-1))G_l^s(\cos\gamma)
\right]. \eqno (B10)
$$
Since this is to be equal to (B7), one obtains two equations for the
coefficients of $e^{ikr}$ and $e^{-ikr}$. It is from these equations
that we arrive at
$$
c_l=2^\frac{d-3}{2}i^le^{i{\delta}_l(s)}(2l+d-2)\frac{\Gamma(s)}
{\sqrt{\pi} k^{{(d-1)}/2}} \eqno (B11)
$$
to finally get
$$
f(\gamma)=\frac{2^{\frac{d-3}{2}}\Gamma(\frac{d-2}{2})
e^{-i\frac{\pi}{4}(d-1)}}{\sqrt{\pi}k^{{(d-1)}/2}}
\sum_{l=0}^{\infty}
\left(l+s\right)\left(e^{2i{\delta}_l(s)}-1\right)G_l^{\frac{d-2}{2}}
\left(\cos\gamma\right). \eqno (B12)
$$
One can easily apply (B12) to the case $d=3$. Indeed, then $G_l^s = P_l$ and
$\Gamma(\half)=\sqrt{\pi}$,  which leads to
$$
f(\gamma)=\frac{1}{2ik}\sum_{l=0}^{\infty}\left(2l+1\right)\left
(e^{2i{\delta}_l}-1\right)P_l(\cos\gamma). \eqno (B13)
$$

\bigskip

\end{document}